\documentstyle[12pt,epsfig]{article}    
    
\def\beq{\begin{equation}}    
\def\eeq{\end{equation}}    
\def\beqa{\begin{eqnarray}}    
\def\eeqa{\end{eqnarray}}    
\def\ba{\begin{array}}    
\def\ea{\end{array}}    
\def\bdm{\begin{displaymath}}    
\def\edm{\end{displaymath}}    
      
\begin{document}    
\input axodraw.sty    
    
\begin{titlepage}    
    
\begin{flushright}    
{\sc UMHEP-438}\\ [.2in]    
{\sc February 24, 1997}\\ [.5in]    
\end{flushright}    
    
\begin{center}    
{\LARGE    
QCD corrections to $B \to J/\psi + {\it anything}$}\\ [.5in]    
{\large Jo\~{a}o M. Soares and Tibor Torma}\\ [.1in]    
{\small    
Department of Physics and Astronomy, University of Massachusetts,\\    
Amherst, MA 01003-4525}\\ [.5in]    
    
{\normalsize \bf Abstract}\\ [.2in]    
\end{center}     
    
{\small    
We calculate the branching ratio for $B \to J/\psi + {\it anything}$, within
the color-singlet approximation for $J/\psi$ production, but including  
perturbative QCD corrections beyond the leading logarithm approximation. Such
higher order corrections are necessary, in order to obtain a result that is
not strongly dependent on the renormalization scale. As in the earlier work of
Bergstr\"om and Ernstr\"om, we use a double expansion in $\alpha_s$ and in the
small ratio of Wilson coefficients $L_0/L_2$, to identify the dominant terms
in the decay amplitude. We complete their work by calculating all the leading
order terms in this double expansion. The predicted branching ratio is then
$B(B \to J/\psi + {\it anything}) = 0.9^{+1.1}_{-0.3} \times 10^{-3}$, which 
is well below the experimental value $B_{exp} = (0.80 \pm 0.08)\%$. This
confirms the suspicion that non-perturbative corrections to the color-singlet  
approximation for $J/\psi$ production in $B$ decays are important.}   
  
\end{titlepage}  
  
\section{Introduction}  
  
The hadronic $B$ decays into charmonium originate in the weak transition $b
\to q c \overline{c}$ (with $q = s$, $d$), followed by the hadronization of
the $c\overline{c}$ pair into the charmonium bound state. They are examples of 
color-suppressed hadronic $B$ decays: at the weak vertex, the $c\overline{c}$ 
pair is not created automatically in a color singlet, and so it will be harder 
for it to hadronize into a charmonium state, rather than into a pair of $D - 
\overline{D}$ mesons. The exclusive decays, such as $B \to K^{(\ast)} J/\psi$, 
and the inclusive decay $B \to J/\psi + {\it anything}$, as well as analogous  
decays into the charmonium states $\psi(2S)$ and $\chi_{c1}$, are well studied 
experimentally. However, color-suppressed hadronic $B$ decays are still not
well understood theoretically.  
  
We will concentrate on the inclusive decay $B \to J/\psi + {\it anything}$. In 
order for the $c\overline{c}$ pair to hadronize into a $J/\psi$, we require
that it forms a color-singlet, with spin $S=1$ and no relative velocity ---
this is the color-singlet mechanism for $J/\psi$ production. Other $c   
\overline{c}$ configurations may also hadronize into $J/\psi$; they appear at  
higher orders in an expansion in the small relative velocity of the $J/\psi$  
constituents \cite{Bodwin}\cite{Ko}. Since they contribute incoherently to the 
$B \to J/\psi + {\it anything}$ decay rate, they can be studied separately.
We are interested in obtaining a reliable prediction for the leading   
color-singlet contribution. Comparing our prediction to the data will   
determine whether other $c\overline{c}$ configurations are indeed important.
  
At first sight, the decay rate in the color-singlet mechanism, with QCD  
corrections included in the leading logarithm approximation, appears to be  
well below the experimental value. However, the result is not satisfactory, as 
it retains a strong dependence on the renormalization scale. To obtain a  
reliable prediction, it is necessary to include higher order QCD corrections. 
This was the subject of the work of Bergstr\"om and Ernstr\"om in Ref.~\cite 
{BandE}. These authors have shown how a clever reorganization of the higher  
order corrections can be used to identify the relevant contributions to the  
decay rate, and how this will eliminate the strong dependence on the  
renormalization scale. Here, we complete their calculation and derive the  
prediction for the $B \to J/\psi + {\it anything}$ decay rate, in the  
color-singlet mechanism.  
  
Up to small corrections of higher order in $\Lambda_{QCD}/m_b$, the inclusive 
decay $B \to J/\psi + {\it anything}$ is described by the corresponding parton 
decay. In the next section, we give the effective weak Hamiltonian for the  
$b$-quark decay, with the next-to-leading order Wilson coefficients of  
Ref.~\cite{Buras}. In the following section, we adopt the program outlined in 
Ref.~\cite{BandE}, and obtain the leading terms in the $b \to q J/\psi$ decay 
amplitude. Finally, we give our numerical results and discuss their  
significance.

\section{The effective weak Hamiltonian for the $b$-quark decay}  
  
The terms of interest in the $\Delta B = 1$ effective weak Hamiltonian  
\beq  
H_{eff} = \frac{G_{F}}{\sqrt{2}} \sum_{q=s,d} V_{cb} V_{cq}^\ast \;  
\left[ \frac{1}{3} C_0(\mu) {\cal O}_1 + 2 C_2(\mu) {\cal O}_8 \right]  
\label{eq:1}  
\eeq  
contain the operators   
\beqa  
{\cal O}_1 & \equiv &   
\overline{c} \gamma_{\mu} (1-\gamma_5) c \;\;   
\overline{q} \gamma^{\mu} (1-\gamma_5)  b \ ,  
\label{eq:2} \\  
{\cal O}_8 & \equiv & \frac{1}{4} \;   
\overline{c} \lambda^a \gamma_{\mu} (1-\gamma_5) c \;\;   
\overline{q} \lambda^a \gamma^{\mu} (1-\gamma_5) b \ ; 
\label{eq:3}  
\eeqa  
they correspond to the $b \to q c \overline{c}$ ($q = s \makebox[2em] {\rm or} 
d$) transition, which occurs at tree level in the weak interaction. The  
subscripts in ${\cal O}_{1,8}$ designate the singlet (1) or octet (8) color  
structure of the $V-A$ currents in those operators. The Wilson coefficients  
$C_{0,2}(\mu)$ include perturbative QCD corrections to the weak vertex.  
They depend on the renormalization scale $\mu$, which effectively separates  
those QCD corrections, from the QCD effects that appear in the matrix elements 
$\langle q J/\psi | {\cal O}_{1,8} | b \rangle$. The $\mu$ dependence of the
matrix elements should cancel that in the Wilson coefficients, so that the  
final result for the decay amplitude is independent of the renormalization  
scale.  
  
In the leading logarithm approximation (LLA), the Wilson coefficients are   
\cite{Buras}  
\beqa  
C_0(\mu) &=& L_0(\mu) = 2 L_+(\mu) - L_-(\mu) \ ,  
\label{eq:4} \\  
C_2(\mu) &=& L_2(\mu) = \frac{1}{2} \; [ L_+(\mu) + L_-(\mu) ]  \ ,  
\label{eq:5}  
\eeqa  
with  
\beq  
L_\pm(\mu) = \left[ \frac{\alpha_s(M_W)}{\alpha_s(\mu)} \right]  
^{\frac{\gamma^{(0)}_\pm}{2 \beta_0}}   
\label{eq:6}  
\eeq  
and  
\beq  
\alpha_s(\mu) = \frac{4 \pi}{\beta_0 \ln (\mu^2/\Lambda_{QCD}^2)} \ ;  
\label{eq:7}
\eeq
whereas at next-to-leading order (NLO), they are \cite{Buras}  
\beqa
C_0(\mu) &=& 2 C_+(\mu) - C_-(\mu) \ ,  
\label{eq:8} \\  
C_2(\mu) &=& \frac{1}{2} \; [ C_+(\mu) + C_-(\mu) ] \ ,  
\label{eq:9}  
\eeqa  
with  
\beqa  
C_\pm(\mu) &=& L_\pm(\mu)
\left[ 1 \pm \frac{\alpha_s(M_W)}{4\pi} \frac{3 \mp 1}{6}   
( 11 \pm \kappa_\pm ) \right. \nonumber \\  
& & \left. + \frac{\alpha_s(\mu)-\alpha_s(M_W)}{4\pi}   
(\gamma^{(0)}_\pm \frac{\beta_1}{2 \beta_0^2} -   
\gamma^{(1)}_\pm \frac{1}{2 \beta_0})  \right]  
\label{eq:10}   
\eeqa  
and  
\beqa  
\alpha_s(\mu) &=& \frac{4 \pi}{\beta_0 \ln (\mu^2/\Lambda_{QCD}^2)}  
\left[ 1 - \frac{\beta_1 \ln [\ln (\mu^2/\Lambda_{QCD}^2)]}  
{\beta_0^2 \ln (\mu^2/\Lambda_{QCD}^2)} \right] \ .   
\label{eq:11}  
\eeqa  
The anomalous dimensions and $\beta$-function coefficients are  
\beqa  
\gamma^{(0)}_\pm = \pm 2 (3 \mp 1) & , &  
\gamma^{(1)}_\pm = \frac{3 \mp 1}{6} \left( -21 \pm \frac{4}{3} \; n_f   
- 2 \beta_0 \kappa_\pm \right) \ ,  
\label{eq:12} \\  
\beta_0 = 11 - \frac{2}{3} \; n_f & , &   
\beta_1 = 102 - \frac{38}{3} \; n_f \ ,  
\label{eq:13}   
\eeqa  
and $n_f = 5$ is the number of active flavors. In the modified minimal   
subtraction ($\overline{MS}$) renormalization scheme, $\Lambda_{QCD} =  
\Lambda^{(5)}_{\overline{MS}} = 209^{+39}_{-33}$ MeV, which corresponds to   
$\alpha_s(M_Z) = 0.118 \pm 0.003$ \cite{PDG96}.  
The quantity $\kappa_\pm = 0$ (NDR), $\mp 4$ (HV) or $\mp 6 - 3$ (DRED) is   
regularization scheme dependent. A similar scheme dependence appears in 
the calculation of the matrix elements of ${\cal O}_{1,8}$, such that the  
final result for the decay amplitude is regularization scheme independent.

\section{The $b \to q J/\psi$ decay amplitude in the $\alpha_s \ - \ L_0/L_2$ 
double expansion}  
  
The amplitude for $b \to q J/\psi$ ($q = s \makebox[2em]{\rm or} d$), when   
the mass of the $q$-quark is neglected, can be parametrized in terms of the
coefficients $g_1$ and $g_2$, that multiply the two possible Lorentz   
structures of the amplitude \cite{Joao1}  
\begin{eqnarray}   
A_{b \to  q J/\psi} &=& - \; \frac{G_F}{\sqrt{2}} \; V_{cb}V_{cq}^\ast \;   
\frac{f_{J/\psi}}{m_{J/\psi}} \;\;  
[ \; g_1 \; m_{J/\psi}^2 \; \overline{u}_q \gamma_\mu (1 - \gamma_5) u_b   
\nonumber \\  
& & + \; g_2 \; m_b \; \overline{u}_q i \sigma_{\mu\nu} p_{J/\psi}^\nu   
(1 + \gamma_5) u_b \; ] \; \varepsilon^{\ast\mu}_{J/\psi} \ .  
\label{eq:14}  
\end{eqnarray}  
The decay rate is then  
\begin{eqnarray}   
\lefteqn{\Gamma(b \rightarrow s J/\psi) + \Gamma(b \rightarrow d J/\psi) =} 
\nonumber\\ 
& & = \frac{G_F^2}{16\pi} \; |V_{cb}|^2 \; 
\left( \frac{f_{J/\psi}}{m_{J/\psi}} \right)^2 m_b^5 (1 - r)^2 \nonumber\\  
& & \times \left[ |g_1|^2 r (1 + 2r) +  |g_2|^2 (2 + r)  
- {\it Re}(g_1 g_2^\ast) 6 r \right] \ ,  
\label{eq:15}   
\end{eqnarray}  
where $r \equiv m_{J/\psi}^2/m_b^2$. The $J/\psi$ decay constant,   
$f_{J/\psi}$, is a non-perturbative parameter that describes the hadronization 
of a color-singlet $c\overline{c}$-pair, with no relative velocity and in a
spin $S=1$ state, into a $J/\psi$ meson. It is defined by 
\beq  
\langle 0 | \overline{c} \gamma^\mu c | J/\psi \rangle = m_{J/\psi}  
\; f_{J/\psi} \; \varepsilon^\mu_{J/\psi} \ ,  
\label{eq:16}  
\eeq  
and it can be measured from the rate for the decay $J/\psi \to e^+ e^-$.   
  
In the LLA, the decay amplitude is obtained from the tree level matrix element 
of the effective weak Hamiltonian in Eq.~\ref{eq:1}. Only the singlet operator 
${\cal O}_1$ contributes (as in Fig.~1a), and one has  
\beqa  
g_1 = \frac{1}{3} \; L_0(\mu) & , & g_2 = 0 \ .  
\label{eq:17}  
\eeqa  
This leads to a LLA branching ratio (see Fig.~3) that depends strongly on the 
renormalization scale $\mu$ (to the point where it can actually vanish, for  
$\mu \simeq 2.5$ GeV!). As pointed out by Bergstr\"om and Ernstr\"om \cite  
{BandE}, this means that higher order contributions to the matrix elements  
of the operators ${\cal O}_{1,8}$, beyond the LLA, need to be considered. 
More specifically, the strong $\mu$ dependence in the LLA amplitude is due  
entirely to the Wilson coefficient $L_0(\mu)$. An expansion of this  
coefficient around some fixed $\mu \sim m_b$ gives  
\beq  
L_0(\mu) = L_0(m_b) + \frac{2}{\pi} \alpha_s(m_b) L_2(m_b)   
\ln (\mu^2/m_b^2) + \dots   
\label{eq:18}  
\eeq  
--- although the $\mu$ dependence only appears at order $\alpha_s$, it is  
important, since $L_0(m_b)/L_2(m_b) \simeq 0.34$ is of similar size as   
$\alpha_s(m_b) \simeq 0.21$. From Eq.~\ref{eq:18}, one can conclude that 
the terms in the matrix elements of ${\cal O}_{1,8}$, which cancel the  
strong $\mu$ dependence that comes from $L_0$, are terms of order $\alpha_s  
L_2$. Such higher order terms in $\alpha_s$ appear at the 1-loop level; they
are absent from the LLA result, where ${\cal O}_{1,8}$ only contribute  
at tree level.  
 
The reason higher order terms in $\alpha_s$ become important is the existence 
of a second small quantity in the calculation --- the ratio $L_0/L_2$. Then, 
$\alpha_s L_2/L_0$ is not small compared to unity, and the expansion in 
$\alpha_s$ fails. Bergstr\"om and Ernstr\"om advocated instead a double 
expansion in both $\alpha_s$ and $L_0/L_2$. Since these are the two  
small quantities in the calculation, this procedure will correctly identify  
the dominant terms in the matrix elements of ${\cal O}_{1,8}$. Those are also 
the terms that are needed to cancel the $\mu$ dependence that comes from the
Wilson coefficients. Here, we will follow this program, and calculate all the 
leading order terms in the double expansion. Inexplicably, this was not done
in Ref.~\cite{BandE}, where the important terms of order $\alpha_s^2 L_2^2$  
in the decay rate were not calculated (and the estimate that was given is 
incorrect).

At leading order in the $\alpha_s - L_0/L_2$ expansion for the decay   
amplitude, there is a tree level contribution from ${\cal O}_1$ (Fig.~1a),   
proportional to $C_0$, and 1-loop contributions from ${\cal O}_8$ (Figs.~1b
and 1c), proportional to $\alpha_s C_2$. From the results for the Wilson   
coefficients in the previous section, and to the order that we are interested, 
\beqa  
C_0(\mu) &=& L_0(\mu) + \alpha_s(\mu) L_2(\mu) \frac{1}{3\pi}   
\left\{ - \kappa + 11 \frac{\alpha_s(M_W)}{\alpha_s(\mu)} \right.   
\nonumber \\  
& & \left. + \frac{\alpha_s(M_W)-\alpha_s(\mu)}{\alpha_s(\mu)}   
\left[ (21 + 4 n_f)\frac{1}{6 \beta_0} - \frac{6 \beta_1}{\beta_0^2} \right]
\right\} \ ,   
\label{eq:19} \\  
C_2(\mu) &=& L_2(\mu)  
\label{eq:20}  
\eeqa  
(where we have replaced the regularization scheme dependent parameter
$\kappa_{\pm}$ by $\kappa = 0$ (NDR), $4$ (HV) or $5$ (DRED)). After a 
lengthy calculation, we obtain, for the parameters $g_{1,2}$ in the decay 
amplitude,  
\beqa  
g_1 &=& \frac{1}{3} L_0(\mu) +  \alpha_s(\mu) L_2(\mu) \frac{1}{9\pi}  
\left\{\kappa^\prime  - \kappa - 6 \ln \left(\frac{\mu^2}{m_b^2}\right)   
 + 11 \frac{\alpha_s(M_W)}{\alpha_s(\mu)}  
\right. \nonumber \\   
& & + \frac{\alpha_s(M_W)-\alpha_s(\mu)}{\alpha_s(\mu)}  
\left[ (21 + 4 n_f)\frac{1}{6 \beta_0} - \frac{6 \beta_1}{\beta_0^2} \right]
\nonumber \\  
& & + \frac{2}{2 - r}  
\left[ - 16 + 7 r + 4 r \ln 2 + \frac{2 r^2}{1 - r} \ln r  
\right. \nonumber \\  
& & \left. \left.   
+ \frac{6 - 6 r + r^2}{2 - r} \ln (1 - r)   
- i \pi \frac{6 - 6 r + r^2}{2 - r} \right] \right\}  
\ , \label{eq:21} \\  
g_2 &=& \alpha_s(\mu) L_2(\mu) \frac{1}{9\pi} \frac{2 r}{2 - r}  
\left[ - 2 + 4 \ln 2 + \frac{2 r}{1 - r} \ln r   
\right. \nonumber \\  
& & \left. + \frac{4 - 3 r}{2 - r} \ln (1 - r)   
- i \pi \frac{4 - 3 r}{2 - r} \right] \ .  
\label{eq:22}  
\eeqa  
  
Both $g_1$ and $g_2$ are free of infrared divergences. The imaginary parts   
correspond to the contribution from the on-shell intermediate state with  
a $c\overline{c}$ color-octet, in the 1-loop diagrams of Fig.~1b. They have
been previously calculated in Ref.~\cite{joao2}. The ultraviolet divergence 
in $g_1$ has been removed using the same $\overline{MS}$ renormalization   
scheme as in the calculation of the Wilson coefficients \cite{Buras}: to  
leading order in the double expansion, the counter-terms that are needed are 
\beq  
H_{c.t.} = \frac{G_{F}}{\sqrt{2}} \sum_{q=s,d} V_{cb} V_{cq}^\ast \;  
\left[ \frac{2}{\varepsilon} - \gamma_E + \ln (4\pi) \right]   
\frac{2}{3} \frac{\alpha_s(\mu)}{\pi} C_2(\mu)   
{\cal O}_1  
\label{eq:22a} 
\eeq 
(for $d = 4 - \varepsilon$ dimensions). The regularization of the divergence
generates the scheme dependent term $\kappa^\prime = -2$ (NDR), $2$ (HV)  
or $3$ (DRED), in the expression for $g_1$ (see the Appendix, for 
more details on how this term is generated). Notice that $\kappa^\prime$  
is such that it cancels the scheme dependence from the Wilson coefficients, 
parametrized by $\kappa$.  
 
As for the dependence of $g_{1,2}$ on the renormalization scale $\mu$, there 
is also an exact cancellation between the $\mu$ dependence of the Wilson  
coefficients and that which originates in the 1-loop matrix elements of the  
${\cal O}_{1,8}$ operators. The latter is shown explicitly in the expressions 
for $g_{1,2}$. The former can be obtained from the expansion of $L_0(\mu)$ 
around $\mu \sim m_b$ in Eq.~\ref{eq:18}, and the analogous results for  
$\alpha_s(\mu)$ and $L_2(\mu)$:  
\beqa  
L_2(\mu) = L_2(m_b) + \dots  & , & \alpha_s(\mu) = \alpha_s(m_b) + \dots \ ,
\label{eq:23}  
\eeqa 
As in Eq.~\ref{eq:18}, the terms not shown are of higher order in the double 
expansion in $\alpha_s - L_0/L_2$. In our final result (see Fig.~3), we have 
chosen to keep the full expressions for $\alpha_s(\mu)$ and $L_{0,2}(\mu)$, 
rather than use the above expansions. This leads to a residual $\mu$  
dependence in the branching ratio, since higher order terms in  
Eqs.~\ref{eq:18} and \ref{eq:23} have not been matched by the corresponding  
higher order contributions to the matrix elements of ${\cal O}_{1,8}$. The  
fact that the residual $\mu$ dependence is small suggests that such higher  
order terms in the matrix elements are negligible; keeping only the leading  
order in the double expansion, as we do in here, is a good approximation. 
 
In order to obtain the branching ratio for $B \to J/\psi + {\it anything}$,  
one must consider, in addition to $b \to q J/\psi$, the contribution to the
inclusive decay from other possible parton processes. The dominant process is 
$b \to q J/\psi g$, as it contributes to the rate at order $\alpha_s L_2^2$
in our double expansion, through the diagrams in Fig.~2. After integrating 
over the three-body phase space, the result for the decay rate is   
\beqa  
\lefteqn{\Gamma(b \to s J/\psi g) + \Gamma(b \to d J/\psi g) =} \nonumber\\
& & = \frac{G_F^2}{162 \pi^2} |V_{cb}|^2 \alpha_s L_2^2  
\left(\frac{f_{J/\psi}}{m_{J/\psi}}\right)^2 m_b^5 \nonumber\\   
& & \times \left[ \frac{1}{6} r (1 - r) (1 + 37r - 8r^2) -  
(1 - 6r) r \ln r \right] \ , 
\label{eq:24}  
\eeqa  
with $r \equiv m_{J/\psi}^2/m_b^2$. Notice that this contribution to the $B 
\to J/\psi + {\it anything}$ decay rate is of lower order in the double 
expansion than the contribution from $b \to q J/\psi$. However, we have 
checked that the three-body phase space suppresses the rate, so that both 
contributions are quantitatively of similar size. This allows us to neglect 
corrections to $b \to q J/\psi g$ and other parton processes such as $b \to q 
J/\psi g g$ or $b \to q J/\psi g q^\prime \overline{q}^\prime$, that appear in
the $B \to J/\psi + {\it anything}$ rate at order $\alpha_s^2 L_2^2$ or 
$\alpha_s L_0 L_2$. They are of the same order as our calculation for the $b 
\to q J/\psi$ rate, but quantitatively smaller because of the phase space  
suppression.

\section{Results}  
  
Using the expressions for the decay rates, in Eqs.~\ref{eq:15} and \ref  
{eq:24}, together with the results of Eqs.~\ref{eq:21} and \ref{eq:22}, we   
obtain our prediction for the $B \to J/\psi + {\it anything}$ branching  
ratio,  
\beq  
B(B \to J/\psi + {\it anything}) = \tau_B \; \sum_{q = s,d} 
[ \; \Gamma(b \to q J/\psi) + \Gamma(b \to q J/\psi g) \; ] \ .  
\label{eq:25}  
\eeq  
In order to eliminate the factor of $m_b^5$ that appears in the decay rates,
and so minimize the uncertainty in our result, we divide the RHS of 
Eq.~\ref{eq:25} by the expression for the inclusive semileptonic 
$B$ decay rate \cite{Cabibbo},  
\beq  
B(B \to X_c e^- \overline{\nu}_e) = \tau_B \; \frac{G_F^2}{192 \pi^3} \;  
|V_{cb}|^2 \; m_b^5 \; f(\frac{m_c}{m_b}) \; h(\frac{m_c}{m_b},m_b) \ ,  
\label{eq:26}  
\eeq  
and multiply it by the experimental result \cite{PDG96}  
\beq  
B(B \to X_c e^- \overline{\nu}_e) = (10.4 \pm 0.4) \% \ .  
\label{eq:26a}  
\eeq  
In Eq.~\ref{eq:26}, 
\beq  
f(z) = 1 - 8 z^2 + 8 z^6 - z^8 - 24 z^4 \ln z \ ,  
\label{eq:27}  
\eeq  
is a phase space factor, and 
\beq  
h(z, \mu)  =  1 - \frac{2 \alpha_s(\mu)}{3 \pi} \tilde{h}(z) \ ,  
\label{eq:28}  
\eeq  
with \cite{Cabibbo} 
\beq  
\tilde{h}(z)  \simeq  (\pi^2 - \frac{31}{4}) \; (1-z)^2 \; + \; \frac{3}{2}  
\ ,  \label{eq:28a}  
\eeq  
is a QCD correction factor. In our quantitative analysis, we take for the 
$b$-quark pole mass $m_b = 4.8 \pm 0.15$ GeV and $m_b - m_c = 3.40$ GeV, as 
in \cite{Greub}. As for the $J/\psi$ decay constant, it can be determined from
$\Gamma(J/\psi \to e^+ e^-) = (5.26 \pm 0.37)$ keV \cite{PDG96}. Within the 
color-singlet approximation (now applied to the $J/\psi$ decay) and including 
perturbative QCD corrections,  
\beq  
\Gamma(J/\psi \to e^+ e^-) = \frac{4 \pi}{3} Q_c^2 \alpha(m_{J/\psi})^2  
\left( \frac{f_{J/\psi}}{m_{J/\psi}} \right)^2 m_{J/\psi}  
\left[ 1 - \frac{16 \alpha_s(m_{J/\psi})}{3 \pi} \right]  
\label{eq:29} \ ;  
\eeq  
with $\alpha(m_{J/\psi}) = 1/133$ and $\alpha_s(m_{J/\psi}) = 0.245$, we find 
$f_{J/\psi} \simeq 515$ MeV. Given the importance of the QCD corrections of 
order $\alpha_s$, it is quite possible that higher order corrections, not 
included in Eq.~\ref{eq:29}, will be significant. Because of this potentially  
large theoretical uncertainty, we choose to factor out the dependence on 
$f_{J/\psi}$ in our final result. 
  
In Fig.\ 3, we show the predicted $B \to J/\psi + {\it anything}$ branching  
ratio as a function of the renormalization scale $\mu$, for $\mu$ ranging
between $m_b/2$ and $2 m_b$. The error band corresponds to the uncertainty in  
$\Lambda^{(5)}_{\overline{MS}}$. Adding to this the uncertainty in $m_b$ and 
that due to the residual $\mu$ dependence, we find  
\beqa
B(B \to J/\psi+\mbox{\it anything}) &=&   
\left( \frac{f_{J/\psi}}{515\ {\rm MeV}} \right)^2   
\times 0.9^{+1.1}_{-0.3} \times 10^{-3} \ .  
\label{eq:30} 
\eeqa
(We have not included the error due to the use of the parton process to 
describe the inclusive decay of the $B$ meson \cite{Savage}. That error  
is small when compared to the sensitivity of our result to the exact value  
of $\Lambda^{(5)}_{\overline{MS}}$ and $m_b$). The central value in 
Eq.~\ref{eq:30} corresponds to $\mu = m_b = 4.8$ GeV and 
$\Lambda^{(5)}_{\overline{MS}} = 209$ MeV. 
 
It is clear that the prediction for the $B \to J/\psi + {\it anything}$ decay
rate, in the color-singlet approximation for $J/\psi$ production and decay,
falls well short of the experimental value $B_{exp} = (0.80 \pm 0.08)\%$.   
Although one might have suspected that this was so, already from the LLA   
prediction, that result could not be trusted because of the strong dependence 
on the renormalization scale (see Fig.\ 3). With a prediction that is much 
more stable in $\mu$, it can now be concluded with certainty that new,  
non-perturbative, contributions to the description of the $J/\psi$ bound-state
are indeed important \cite{Bodwin}. A calculation of the $B \to J/\psi + 
{\it anything}$ decay rate which attempts to include such contributions is 
given in Ref.~\cite{Ko}. Our result can be used \cite{fpsi} to improve the 
color-singlet part of the $B \to J/\psi + {\it anything}$ decay rate that 
appears in there.

\section*{}  
 
We wish to thank John Donoghue and Alexey Petrov for numerous discussions.  
This work was supported, in part, with a grant from the Department of Energy
(T.T.) and a grant from the National Science Foundation (J.M.S.).

\section*{Appendix} 
 
The origin of the scheme dependent term $\kappa^\prime$, in the expression  
for $g_1$ (see Eq.~\ref{eq:21}), deserves a brief comment. We summarize the  
more detailed discussion of Ref.~\cite{BurasWeisz} (notice that our result  
for $\kappa^\prime$ can be recovered from the more general case presented 
there). 
 
The $\kappa^\prime$ term in $g_1$ originates from the regularization of the  
divergence in the un-renormalized 1-loop $b \to q J/\psi$ amplitude. That  
divergence is of the form  
\beqa 
A_{div} &=& - \frac{G_{F}}{\sqrt{2}} V_{cb} V_{cq}^\ast \; 
C_2 \alpha_s \frac{1}{36\pi} \nonumber \\  
& & \times \left[ \frac{2}{\varepsilon} - \gamma_E + \ln (4\pi) \right] 
\langle q J/\psi | \Omega | b \rangle \ ,  
\label{eq:A1}  
\eeqa  
with 
\beqa 
\Omega &=& \overline{q} [ \gamma_{\mu} (1 - \gamma_5) \gamma_{\alpha}  
\gamma_{\nu} - \gamma_{\nu} \gamma_{\alpha} \gamma_{\mu} (1-\gamma_5) ] b  
\nonumber\\ 
& & \times \overline{c} [ \gamma^{\nu} \gamma^{\alpha} \gamma^{\mu}  
(1-\gamma_5) - \gamma^{\mu} (1 - \gamma_5) \gamma^{\alpha}  
\gamma^{\nu} ] c \ . 
\label{eq:A1a} 
\eeqa 
In $d = 4$ dimensions, ${\cal O}_{1,8}$ form a complete basis under QCD  
corrections, and $\Omega$ is proportional to ${\cal O}_1$: $\Omega =  
- 24 \; {\cal O}_1$. In $d = 4 - \varepsilon$ dimensions, however, the  
${\cal O}_{1,8}$ basis must be extended to include evanescent  
operators $E$, and 
\beq 
\Omega = a_1(\varepsilon) {\cal O}_1 + E \ . 
\label{eq:A2}  
\eeq
The evanescent operators $E$ do not exist in 4 dimensions, and they do 
not contribute to the physical amplitude. On the other hand, the terms  
of order $\varepsilon$ in $a_1(\varepsilon)$ will contribute to the 
finite part of the $b \to q J/\psi$ amplitude, when inserted in  
Eq.~\ref{eq:A1a}. In order to determine these terms, we must completely 
fix the regularization scheme by giving the form of the evanescent 
operators E (and the same form for these operators must be used in the  
calculation of the Wilson coefficients). One way to do this is to define
the evanescent operators by the condition \cite{BurasWeisz}
\beq  
E_{\alpha\beta,\gamma\delta} \Gamma_{\beta\gamma} \Gamma_{\delta\alpha}  
= 0 \ ,
\label{eq:A3}
\eeq
where the two pairs of fermion fields were removed from $E$, and each one
was replaced by $\Gamma \equiv \gamma_\mu (1+\gamma_5)$. Applying this 
operation to both sides of Eq.~\ref{eq:A2}, we obtain  
\beqa  
a_1(\varepsilon) &=& \Omega_{\alpha\beta,\gamma\delta} \Gamma_{\beta\gamma} 
\Gamma_{\delta\alpha} \; 
\left( {\cal O}_{1\;\alpha\beta,\gamma\delta} \Gamma_{\beta\gamma} 
\Gamma_{\delta\alpha}  \right)^{-1} 
\nonumber \\
&=& - 12 (2 - \kappa^\prime \varepsilon + {\cal O}(\varepsilon^2) ) \ . 
\label{eq:A4}  
\eeqa 
This is the origin of the scheme dependent $\kappa^\prime$ term in the  
expression for $g_1$ of Eq.~\ref{eq:21}.

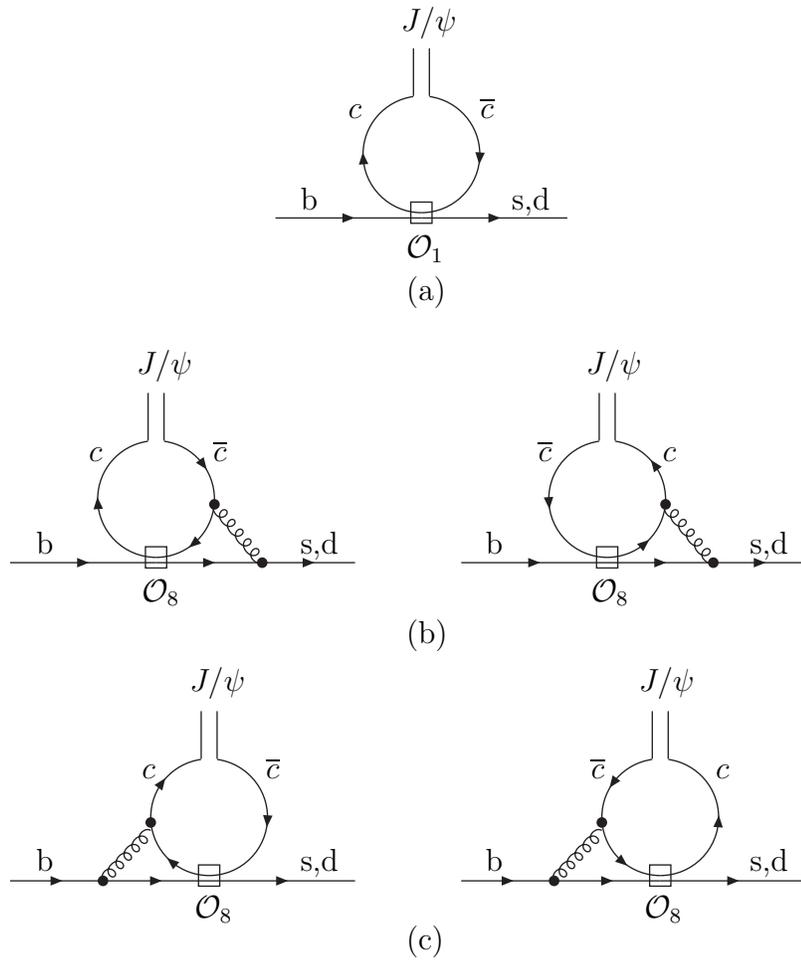
\begin{figure}  
\begin{picture}(315,360)(0,0)  
\ArrowLine(130,266)(185,266)  
\put(140,270){b}  
\ArrowLine(185,266)(240,266)  
\put(220,270){s,d}  
\EBox(181,264)(189,272)  
\ArrowArcn(185,290)(22,82,270)  
\put(158,305){$c$}  
\ArrowArcn(185,290)(22,270,98)  
\put(208,305){$\overline{c}$} 
\Line(182,312)(182,330)  
\Line(188,312)(188,330)  
\put(180,252){${\cal O}_1$}  
\put(180,236){(a)}  
\put(178,338){$J/\psi$}  
\ArrowLine(30,136)(85,136)  
\put(40,140){b}  
\ArrowLine(85,136)(125,136)  
\ArrowLine(125,136)(160,136)  
\Vertex(125,136) 2  
\Vertex(107,158) 2  
\Gluon(125,136)(107,158) 2 5  
\put(140,140){s,d}  
\EBox(81,134)(89,142)  
\ArrowArcn(85,160)(22,350,270)  
\ArrowArcn(85,160)(22,82,350)  
\put(60,175){$c$}  
\ArrowArcn(85,160)(22,270,98)  
\put(107,175){$\overline{c}$}  
\Line(82,182)(82,200)  
\Line(88,182)(88,200)  
\put(80,122){${\cal O}_8$}  
\put(180,106){(b)} 
\put(78,208){$J/\psi$}  
\ArrowLine(200,136)(255,136)  
\put(210,140){b}  
\ArrowLine(255,136)(295,136)  
\ArrowLine(295,136)(330,136)  
\Vertex(295,136) 2  
\Vertex(277,158) 2  
\Gluon(295,136)(277,158) 2 5  
\put(310,140){s,d}  
\EBox(251,134)(259,142)  
\ArrowArc(255,160)(22,270,350)  
\ArrowArc(255,160)(22,350,82)  
\put(230,175){$\overline{c}$}  
\ArrowArc(255,160)(22,98,270)  
\put(277,175){$c$}  
\Line(252,182)(252,200)  
\Line(258,182)(258,200)  
\put(250,122){${\cal O}_8$}  
\put(248,208){$J/\psi$}  
\ArrowLine(30,16)(65,16) 
\ArrowLine(65,16)(105,16)  
\put(40,20){b}  
\ArrowLine(105,16)(160,16) 
\Vertex(65,16) 2  
\Vertex(83,38) 2  
\Gluon(65,16)(83,35) 2 5  
\put(140,20){s,d}  
\EBox(101,14)(109,22)  
\ArrowArcn(105,40)(22,190,98)  
\ArrowArcn(105,40)(22,270,190)  
\put(80,55){$c$}  
\ArrowArcn(105,40)(22,82,270)  
\put(127,55){$\overline{c}$}  
\Line(102,62)(102,80) 
\Line(108,62)(108,80)  
\put(100,2){${\cal O}_8$}  
\put(180,-10){(c)} 
\put(98,88){$J/\psi$}  
\ArrowLine(200,16)(235,16)  
\ArrowLine(235,16)(275,16)  
\put(210,20){b}  
\ArrowLine(275,16)(330,16)  
\Vertex(235,16) 2  
\Vertex(253,38) 2  
\Gluon(235,16)(253,35) 2 5  
\put(310,20){s,d}  
\EBox(271,14)(279,22)  
\ArrowArc(275,40)(22,98,190)  
\ArrowArc(275,40)(22,190,270)  
\put(250,55){$\overline{c}$}  
\ArrowArc(275,40)(22,270,82)  
\put(297,55){$c$}  
\Line(272,62)(272,80)  
\Line(278,62)(278,80)  
\put(270,2){${\cal O}_8$} 
\put(268,88){$J/\psi$}  
\end{picture} 
\caption{Feynman graphs contributing to $B \to q J/\psi$ ($q = s$, $d$),  
to lowest order in the $\alpha_s - L_0/L_2$ double expansion.}  
\end{figure}  
  
\begin{figure} 
\begin{center}  
\begin{picture}(315,100)(0,0)  
\ArrowLine(30,16)(85,16)  
\put(40,20){b}  
\ArrowLine(85,16)(140,16)  
\Vertex(107,38) 2  
\Gluon(150,50)(107,38) 2 5  
\put(133,53){g}  
\put(120,20){s,d}  
\EBox(81,14)(89,22)  
\ArrowArcn(85,40)(22,350,270)  
\ArrowArcn(85,40)(22,82,350)  
\put(60,55){$c$}  
\ArrowArcn(85,40)(22,270,98)  
\put(107,55){$\overline{c}$}  
\Line(82,62)(82,80)  
\Line(88,62)(88,80)  
\put(80,2){${\cal O}_8$}  
\put(78,88){$J/\psi$}  
\ArrowLine(200,16)(255,16)  
\put(210,20){b}  
\ArrowLine(255,16)(310,16)  
\Vertex(277,38) 2  
\Gluon(320,50)(277,38) 2 5  
\put(303,53){g}  
\put(290,20){s,d}  
\EBox(251,14)(259,22)  
\ArrowArc(255,40)(22,270,350)  
\ArrowArc(255,40)(22,350,82)  
\put(230,55){$\overline{c}$}  
\ArrowArc(255,40)(22,98,270)  
\put(277,55){$c$}  
\Line(252,62)(252,80)  
\Line(258,62)(258,80)  
\put(250,2){${\cal O}_8$}  
\put(248,88){$J/\psi$}  
\end{picture}  
\end{center}  
\caption{Feynman graphs contributing to $B \to q J/\psi g$  
($q = s$, $d$), to lowest order in the $\alpha_s - L_0/L_2$  
double expansion.} 
\end{figure}
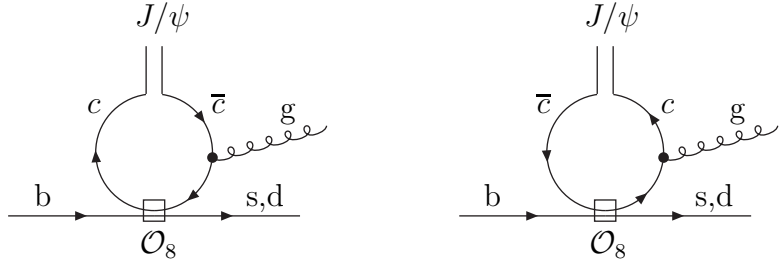

\begin{figure}  
\unitlength0.05in  
\begin{picture}(100,75)  
\put(0,0){\makebox(100,75)  
{\epsfig{figure=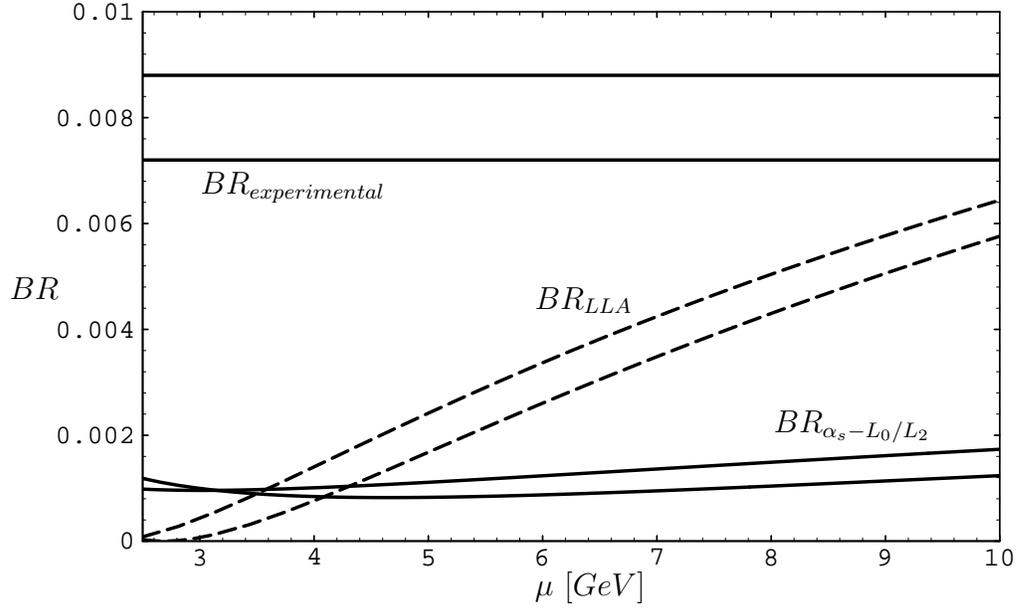,width=5in}}} 
\put(15,47){$BR_{experimental}$}  
\put(50,5){$\mu\ [GeV]$}  
\put(50,35){$BR_{LLA}$}  
\put(75,22){$BR_{\alpha_s-L_0/L_2}$} 
\put(-5,36){$BR$}  
\end{picture}  
\caption{The $B \to J/\psi + anything$ branching ratio as a function of  
the renormalization scale $\mu$. The theoretical curves, in LLA and in the  
$\alpha_s - L_0/L_2$ double expansion, correspond to $f_{J/\psi} = 515$ MeV,  
$m_b = 4.8$ GeV and $\Lambda^{(5)}_{\overline{MS}} = 209^{+39}_{-33}$ MeV.} 
\end{figure}

\end{document}